# Education Stats Made Visible: Helping School District Managers Write Better Three-Year Plans

Steve Rees

School Wise Press

San Francisco, CA

Steve.rees@schoolwisepress.com

Mary Perry

K12 Measures

San Jose, CA

Mlperry2@gmail.com

## ABSTRACT

**1.** <u>Problem:</u>  School district leaders in California are awash in a sea of data, but are often unable to find it, query it, or relate it with other data. Districts are islands, leaving district managers able to see only their own data. A state education agency gathers and republishes districts' data, but doesn't clean it.  New laws now require them to plan in public, set goals, assign resources, and project benefits for specific groups of kids. The evidence base they need remains out of reach.  <u>Solution:</u>  Our team was commissioned by two county offices of education to design a comparative methodology and data visualization environment to help district managers write these three-year plans. Ten year of data spanning finance, staffing, students, discipline, course offerings and assessments were acquired, QA'd and integrated. A Tableau development team built 50+ pages of leaderboards and scatterplots. A documentation team wrote training materials. Each district in this county received a custom workbook which framed its district's vital signs within a context of 15 districts whose students were most like their own.  This presentation shares the work and our evaluation.

## 2. INTRODUCTION

Despite more than 14 years of calls, pleas and admonitions to educators to do "data-driven decision-making," this accountability era has conditioned education leaders to consider data to be a cost at best, and at worst, an instrument of punishment. Overcoming this has resulted in a literature replete with calls for creating a "data driven culture." But the barriers to analytic methods of inquiry in education are more deep seated than this. They are rooted in a profession which defends the privacy of a classroom, where management of teachers is considered an intrusion. It is also a profession dominated by practitioners, whose schools of education have been largely resistant to teaching even the measurement of learning (Popham) (Unlearned Lessons, Harvard Education Press, 2009) as well as the measurement of organizational effectiveness (Jim Guthrie, 2006), or the economics of cost-benefit analysis (Levin, 2011). James Popham, in fact, in his book <u>Unlearned Lessons</u>, titled one chapter "Abysmal Assessment Literacy." Unlike education, the medical profession enjoys a healthy debate between those who advocate for evidence-based medicine, and those who champion the superior wisdom of expertise distilled from practitioners' insight.

In education, the analysts are vastly outnumbered.

Our challenge was to assist education managers in California with their three year plans by building for them a body of evidence of organizational vital signs. From this body of evidence they were to select goals, set progress markers for each of three years for each of them, and defend their decisions in public meetings. Because planning teams were largely inexperienced in planning and analysis, we believed we could only succeed with a visual presentation of patterns inherent in the data. Our hypothesis was that by combining a comparative methodology with a visual mode of analysis, that we could both broaden the number of participants who put our service to work, improve the accuracy of the inferences they derived, and deepen the level of questions they dared bring to the table.

Our team had 14 years of prior experience with California education data, and had published annual reports during this time for over 240 of the state's districts. As a result, we were familiar with the human barriers to adoption, and the frustration that resulted from the state agency's decision to not provide data quality assurance, analytic support, or technical tools. We had studied the history of failed efforts to field data coaching, and the weaknesses of data literacy curricula in schools of education (Mandinach, 2012). Our advantages included our awareness of prior failures, and our deep domain knowledge of K12 data itself.

The customary approach of education managers is simply to make sure that data is being used at all. The lack of regard for analytic craft skill is evident in the absence of regard for the quality of practitioners' inferential judgment. This was the topic of a dissertation and subsequent presentations by Jenny Rankin (2013) who reported that California teachers made incorrect inferences from test score data 48 percent of the time.

We fielded the first installment of this work in December 2015, and the final third installment in April 2016. Thirty-four districts in Monterey and Santa Cruz counties in California, were the beneficiaries of the work, and were supported in their use of it at three workshops held at the county office. This summer we are evaluating the work, combining three methods: survey, interview, and before-vs-after comparisons of plans. Our intent is to unveil our findings at the Bloomberg Data-for-Good conference on Sept. 25, 2016.

## 3. WOULD (VISUAL INTERROGATION) + (COMPARATIVE FRAMING) = IMPROVED COMPREHENSION?

In the winter and spring of 2016, district planners in Monterey and Santa Cruz counties were able to shoulder the challenge of writing their local control accountability plans equipped with an unusual degree of analytic support. Through the auspices of these two county offices of education, 34 districts received a ten-year history of their districts' key data, in a visual form that encouraged active interrogation.

District planning teams encountered a custom Tableau workbook which featured leaderboards, scatterplots, multi-panel trend lines, and rate-of-change representations of this data. Because the end-users were not accustomed to the visual vocabulary or the interactive nature of these visualizations, we paid special attention to training materials, embedded documentation, and to workshop presentations. If they wished, they could avoid reckoning with any numbers in determining where their district led or lagged spanning over fifty factors in these seven categories: finance, staffing, discipline, student demography, course offerings, assessments, and graduation and dropout rates.

The burden of learning a new design vocabulary, and abandoning the familiar zone of bar charts and trend lines, proved to be no barrier to adoption. In fact, the novelty of having new ways to see patterns in the data, easily outweighed any disadvantage the intellectual effort of learning new visual expressions might have imposed.

Avoiding encounters with numbers was considered by most participants to be a benefit. Many in private discussions admitted to a certain fatigue after years of having to comprehend and interpret vast amounts of table based data. The absence of interpretive assistance from almost all data delivery systems in the school world, combined with the increasing volume of data that managers of schools have to digest, was the source of this fatigue. The visual mode of interrogation, combined with the opportunity to actively filter the displayed data (e.g., selecting year, student subgroup, district funding) proved to appear inviting, based on their active attention throughout three workshops lasting up to two hours.

This method of visual inquiry also framed each district within a set of 15 districts whose students most closely resembled their own. In a state with 956 school districts, discovering that there are 15 districts of the same scale, that serve identical grade ranges, whose students' parents have similar education levels, and whose students have similar levels of students learning English and qualifying for subsidized lunches, is in itself a valuable finding. In our discussions with participants, the thing many valued the most was, in fact, the identification of these most similar districts.

## 4. EVALUATION METHODS AND BARRIERS TO ANALYTIC DECISION-MAKING IN CALIFORNIA

We tested three hypotheses in this experiment: (1) would this body of evidence improve the process of plan creation itself, (2) would it result in better plans, and (3) would it improve the communication of plans among district staff and to stakeholders attending public meetings.

The research that accompanies this pilot is an evaluation by the team that created the work, called an LCAP Data Almanac. Surveys of users, combined with interviews, gathered evidence of impact. Self-reported changes by LCAP planners comprise one body of evidence in the evaluation. "Before" and "after" comparisons of plans themselves comprise the rest of the evaluation. This session is the first sharing of our study.

Our evaluation's focus is on three dimensions: (1) to what degree does an interactive environment where quantitative relations are visualized (and numbers avoided) engage the LCAP planning members to be more deliberate in their setting of goals; (2) to what extent does framing a district's vital signs within a context of 15 districts whose students most resemble their own improve the LCAP planning team's decisions about priorities and the setting of goals; (3) does the presentation of a ten year evidence base help or hinder LCAP planning teams in setting goals and determining priorities. The evaluation began in July and is still in progress. We expect to write up results by late October.

### 4.1. BARRIERS TO ANALYTIC INTERROGATION OF SCHOOL DATA IN CALIFORNIA

We are appreciative of the differences among the states in the ways that their laws, policies and traditions vary in the governance of schooling and the management of school districts. We are therefore containing our observations to the state we know well: California. The external appraisal of the condition of education data stewardship and use in California comes from the Data Quality Campaign, a national organization known for its advocacy of modern-minded, enlightened approaches to putting education data to work in service to better decision-making by leaders. Paige Kowalski, in an email correspondence, referred to California being "ten years behind the rest of the country" in its policies and practices. With that in mind, here are the barriers we identified, and our approaches to clearing those barriers off the path that education managers are walking.

### 4.2. STATE DATA CONTAINS ERRORS

The California Dept. of Education (CDE) believes that its responsibility is to disseminate district provided data. If they receive data they believe to be incorrect, they do not correct, but rather give districts a fixed amount of time to correct it. Once that deadline is past, districts the data submission is considered final and they will not revise it. We have identified noise in the signals of every data release from the CDE, and rigorously clean data when possible, or suppress untenable outliers altogether rather than publish it.

### 4.3. STATE DATA MIXES DISSIMILAR ENTITIES TOGETHER

School districts are commingled with charter schools in the master data set of the CDE. Charter schools are associated with the districts that authorize them, rather than with the organization that runs them. This results in a variety of downstream errors. One is that districts are given both the credit and blame for statistical evidence of results that occur in schools they do not govern. Another is that charter management organizations, which often create and manage schools in a variety of districts, are not recognized as the "parent" for the school of record. We correct for these errors by removing direct-funded charter schools' students, teachers and test scores from the districts that have authorized them. We reassociate those schools with the charter management organization that does, indeed, operate them.

## 4.4. STATE EDUCATION AGENCY CREATES RATIOS THAT MULTIPLY THOSE ERRORS

With one in ten California schools now a charter school, this misattribution of charter schools to districts results in ratio analyses that can be costly. The worst of these consequences is the calculation of the cost-per-student for districts. Because the revenue of districts which have authorized charter schools is reported in full for district-governed and charter-governed schools, and because the count of students in those districts excludes charter-school students, these districts appear to be receiving more revenue per student than they do, in fact, receive. We correct for this error.

## 4.5. STATE EDUCATION AGENCY SILOS DATA ELEMENTS MAKING IT DIFFICULT TO EXPLORE RELATIONS

The separation of data about staffing levels, course offerings, test scores, teachers and funding make correlating these data nearly impossible for all but the most skilled researchers and scholars. We have made it possible for users to explore these correlations by building scatterplots designed expressly for this purpose.

## 4.6. STATE EDUCATION AGENCY PROVIDES DISTRICTS WITH NO COMPARATIVE CONTEXT

Districts attempting to plan have only their own data to work with. These purely descriptive statistics enable them to compare their own vital signs in relation to their own history, and to state and county averages. We build districts a comparative context: the 15 districts whose students are most like their own in demographic composition.

## 5. CONCLUSION

We are eager to enable more modern-minded education leaders to harness the power of education data to plan with their eyes wide open. This experiment will reveal if our premises are correct: that education data's meaning can be revealed with greater success to a wider audience of users using visual analytic methods; that its users can enjoy a higher rate of correct inferences if the data is presented in a comparative framework with a small set of closely matched districts; that other staff and members of the public can make sense of the plans of school districts more easily and more accurately if assisted by intelligently built visual interpretations.

The policy consequences of an independent third-party investment in these public benefits could support a Gov 2.0 approach, which would guide state agency leaders to invest in more careful stewardship of the district provided data assets, improve in the quality of the data, and in creating application program interfaces (APIs) that enable others to more effectively publish the data. This policy question may percolate slowly, and require a change in the governor's office before it gets a fair test.

## 6. ACKNOWLEDGEMENTS

The authors value the contributions of several advisors over the past year. These include Doug McRae and Howard Herl, psychometricians whose technical guidance has proven to be invaluable. Thanks, too, to Prof. Henry Levin, Teachers College, Columbia University, for his good advice on research methods and economic questions. Appreciation is also due to his team at the Center for Benefit-Cost Studies in Education at Columbia University. Other team members include Greg Smith of Data Works, Portland, Oregon, and Jonathan McIlroy, who is staffing curriculum development and evaluation during the summer of 2016.

# Sample pages from the Tableau workbook for School District Planners

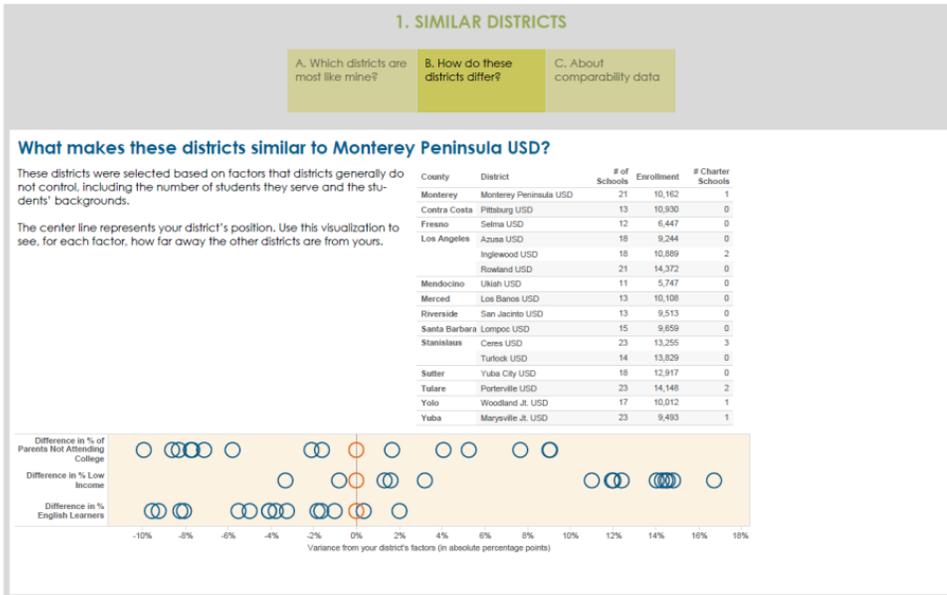

Page showing the degree to which each of 15 districts is similar to the client district, using a bubble chart. Client can mouse over any bubble, and see that district on all three rows.

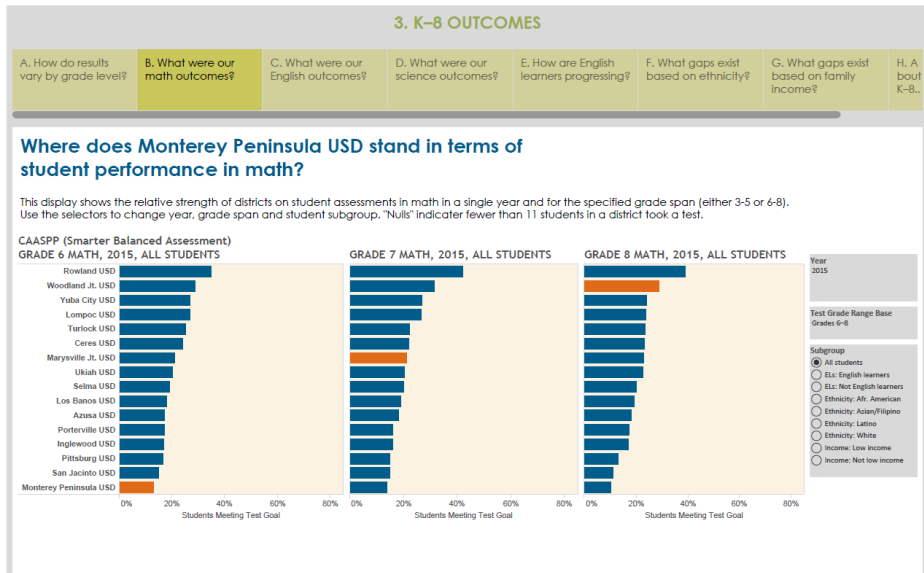

The client district's students' math scores for grades 6, 7 and 8 are visible as orange bars on this leaderboard. Client can select the year to be displayed and the subgroup of students to display.
44